\documentclass[12pt]{article}

\input epsf
\usepackage{epsfig}
\usepackage{graphicx,epsf,amsmath,amsfonts,amssymb,amsbsy}

\title{ Charge-exchange reactions from the standpoint \\
of the parton model }
\author{M. L. Nekrasov \\
{\small\it 
SRC Institute for High Energy Physics of NRC ``Kurchatov
Institute'',} \\
{\small\it Protvino 142281, Russia}}

\date{}
\begin{document}
\maketitle

\begin{abstract}
Using simple arguments, we show that charge-exchange reactions at high
energies go through the hard scattering of fast quarks. On this basis we
describe $\pi^- p \to M^0 n$ and $K^- p \to M^0 \Lambda$, $M^0 =\pi^0,
\eta, \eta'$, in a combined approach which defines hard contributions in
the parton model and soft ones in Regge phenomenology. The disappearance
of a dip according to recent GAMS-4$\pi$ data in the differential
cross-section $K^- p \to \eta\Lambda$ at $|t| \approx$ 0.4--0.5
(GeV$\!$/c)$^2$ at transition to relatively high momenta, is explained
as a manifestation of a mode change of summation of hard contributions
from coherent to incoherent. Other manifestations of the mentioned mode
change are discussed. Constraints on the $\eta$--$\eta'$ mixing and
gluonium admixture in $\eta'$ are obtained.
\end{abstract}

\section{Introduction}\label{sec1}

In accordance with modern ideas, high-energy interactions between
hadrons  occur through elementary events of partons scattering
\cite{Feynman}. The greatest contributions arise from the scattering of
partons with small relative momenta. In the center-of-momentum frame 
they are slow. Such partons always present in hadrons due to quantum
fluctuations, the splitting of fast partons into slow ones followed by
recombination again into the fast partons. Since scattering of partons
typically interrupts the parent fluctuations, the scattering of slow
partons results in the release of a large number of uncorrelated
partons. Being uncorrelated, they can no longer recombine, and because
of dispersion in space form new hadrons, leading thus to a multiple
hadron production. (See {\it e.g.}~discussion in \cite{Gribov}.) An
exception in this picture is the small-angle scattering which can occur
without interruption of the fluctuations. In this case elastic or
quasi-elastic processes are realized.

The charge-exchange reactions represent another type of processes. A
distinguishing feature of them is that they necessarily include 
charge-exchange scattering of charged partons---the valence or sea
quarks, which means changing their type in the composition of hadrons. A
scattering of this kind of slow partons, even on small angles, leads to
destruction of parent fluctuations followed by a multiple hadron
production. So the exclusive charge-exchange reactions go through the
scattering of fast quarks standing in the beginning of quantum
fluctuations. In this case the uncorrelated partons that arise due to
interruption of the fluctuations are fast, as well, and therefore can be
captured by the flying away partons clusters which form new hadrons.

In this way the flavor content of final states in the charge-exchange
reactions is determined by scattering of fast quarks. The appropriate
subprocesses are hard, as they involve high internal virtualities in the
$s$ or $u$ channel (see below). Simultaneously, soft subprocesses
collect partons in the final states and through series of scatterings on
small angles form main contributions to the cross-section. In relation
to the fast hard subprocesses the soft ones, as conventionally assumed,
play a role of background. However they can also be responsible for the
formation of mode of summation of hard contributions. We mean that
intermediate contributions may be coherent or incoherent (in limiting
cases), and this determines what one should sum up at calculating the
cross-sections, the amplitudes or probabilities.  

At first glance, one can expect a coherence of the hard contributions
because unlike a deep inelastic scattering the hard-scattered quarks in
the charge-exchange reactions do not leave the interaction region.
Consequently the hadronization occurs without the fragmentation phase
which is stochastic in its nature and independent from the scattering.
However, the uncorrelated partons that arise due to destructions of
quantum fluctuations form an environment (like a parton gas) which
softly interacts with coherent partons clusters. Moreover, this
environment is completely absorbed by the clusters in the case of
exclusive reactions. This promotes destruction of the coherence in the
intermediate states.

Currently there is no clarity which option is actually realized.
Nevertheless, we can identify conditions that facilitate the loss of the
coherence. Really, the probability for uncorrelated partons be captured
should be increasing with increasing duration of interaction between
hadrons. So the number of absorbed uncorrelated partons should be
increasing with the increasing duration, and this should increase
destruction of the coherence. Simultaneously, the duration of
interactions between hadrons is increasing with increasing the energy of
the collisions. (This is determined by preparation of slow partons,
their soft scatterings, and recombination of quantum fluctuations, see
discussion in \cite{Gribov}.) So, we can expect a loss of the coherence
in the intermediate states with increasing the energy. However, how high
must be energy in order that the coherence should be lost, and what
observable effects as a result should become apparent are issues that
remain unclear.  

On the other hand, among the data on the charge-exchange reactions there
are phenomena that emerge with increasing the energy and which have no
explanation. The most remarkable one is a radical change of the behavior
of the differential cross-section $K^{-}p \to \eta \Lambda$ depending on
the transfer. Namely, there are series of experiments carried out at
CERN hydrogen bubble chamber with the $K^-$ momenta from 3.13 GeV$\!$/c
to 8.25 GeV$\!$/c, in which a pronounced dip was detected in the above
cross-section at $-t \approx$ 0.4--0.5 (GeV$\!$/c)$^2$
\cite{Mason,Moscoso,Marzano,Harran}. The presence of a dip was explained
\cite{Martin} by the dominance in this process of the vector-exchange
trajectory and simultaneously its zeroing by signature factor in the
above mentioned $t$-region. However, in accordance with data \cite{GAMS}
obtained at 32.5 GeV$\!$/c with GAMS-$4\pi$ spectrometer, there is no
dip in this region. Moreover, the behavior of the differential
cross-section is almost purely exponential, at least, up to $t=0.7$
(GeV$\!$/c)$^2$ \cite{GAMS}. Why mechanism \cite{Martin} stops working
with the increasing energy is unclear. Another incomprehensible effect
is the change with increasing the energy of the slope in logarithmic
scale of the ratio of the differential cross-sections $\pi^{-}p \to
\eta' n$ and $\pi^{-}p \to \eta n$. A hint of this effect was originally
observed in \cite{Stanton} at comparing its results with \cite{NICE1},
but the large errors did not allow to make a firm conclusion. However at
comparing \cite{Stanton} with data \cite{GAMS} obtained at higher
statistics, the effect gets confirmation (see below). 

In this paper we study the above mentioned charge-exchange reactions.
We mainly consider a restricted region of the transfer in which the
above effects were detected. We propose a combined approach, which
defines hard contributions in the parton model while soft ones in Regge
phenomenology. We carry out calculations in two cases, with preservation
and complete destruction of the coherence of hard contributions. On the
basis of comparison of the results with data, we make a conclusion about
the real mode of summation of hard contributions. Simultaneously we
study other properties of the above reactions and extract an information
about the mixing of light pseudoscalar states. 

The paper is organized as follows. In the next section we analyze 
subprocesses with fast quarks related to the above reactions. Based
on this analysis, in sect.~\ref{sec3} we define the differential
cross-sections and carry out a comparison of the results with the data.
Section \ref{sec4} discusses the issue of gluonium admixture in $\eta'$.
In sect.~\ref{sec5} we discuss the results and make conclusions.

\section{Analysis of hard subprocesses}\label{sec2}

\begin{figure}[t]
\hbox{ \hspace*{60pt}
       \epsfxsize=0.7\textwidth \epsfbox{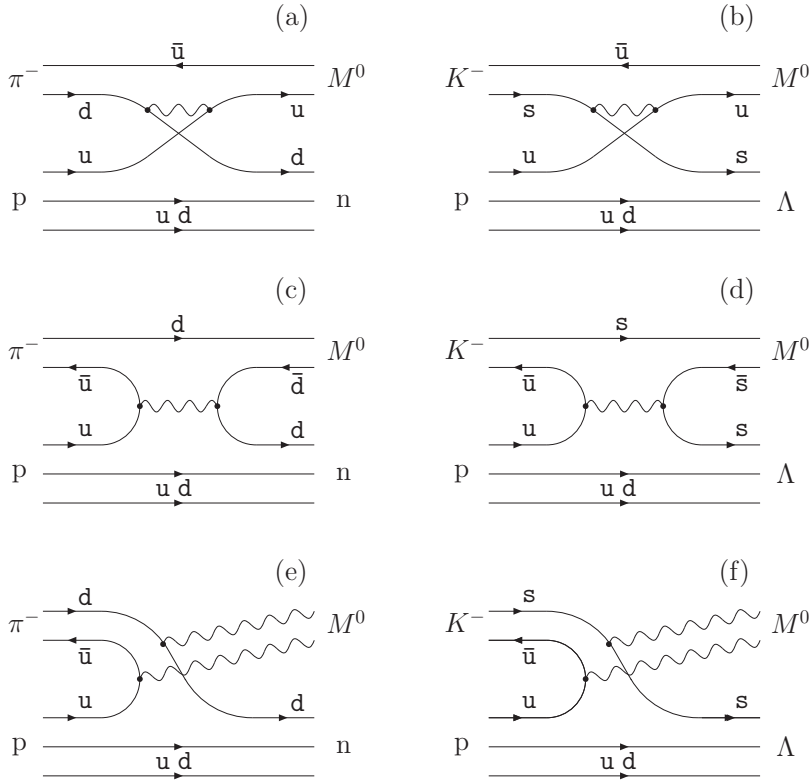}}
\caption{\small Hard subprocesses with valence quarks scattering in
$\pi^{-}\,p \to M^0\,n$ (a,c,e) and $K^{-}\,p \to M^0\Lambda$
(b,d,f), $M^0\!=\! \pi^0, \eta, \eta'$. Solid lines mean valence quarks,
waved lines mean virtual gluons.} \label{Fig1} 
\end{figure}

We consider charge-exchange reactions in high-energy collisions of
$\pi^-$ and $K^-$ with protons followed by production of one-particle
pseudo-scalar states $\pi^{0}$, $\eta$, $\eta'$, and appropriate baryon
states, neutron $n$ or hyperon $\Lambda$. Primarily we restrict
ourselves to region of not too large transfer. As noted in the
introduction, the mentioned reactions occur through subprocesses with
the charge-exchange of fast quarks. Independently of their origin,
either valence or sea, there are three types of relevant subprocesses in
the leading order in the coupling constant. They are with the quark
exchange (E), quark annihilation (A), and with both types of
contributions with simultaneous production of colorless pair of gluons
(G). In fig. \ref{Fig1} we show appropriate diagrams in the case of
valence quark scattering. All diagrams imply hard subprocesses as they
involve high virtuality of intermediate gluons. Really, in fig.
\ref{Fig1}(a,b) the incident quark joins the target, knocking out
another quark with hard momentum. In this case the hard momentum is
transferred via a virtual gluon in the $u$-channel (high negative
virtuality). In fig.\ref{Fig1}(c,d) a large energy is released due to
annihilation of quarks in the $s$-channel, followed by production of
quark-antiquark pair (high positive virtuality). In fig.\ref {Fig1}(e,f)
both processes occur, with the difference that the energy and momentum
are transferred in two gluons. Actually these gluons are virtual and to
be transformed through soft exchanges into valence gluons or valence
quark-antiquark pair. At this stage we do not distinguish between these
two options.

It is worth mentioning that the diagrams in fig.~\ref{Fig1} appear not
only in the leading order in the QCD coupling constant, but in the
common order in the $N_{c}^{-1}\,$-expansion, as well, where $N_{c}$ is
the number of colors. For the diagrams of upper two rows this is obvious
in view of the fact that they are planar. (The diagrams of the first row
can be transformed to explicitly planar form by means of 180$^0$
rotation of their upper halves, see discussion in~\cite{tHooft}.) The
diagrams in the lower row are not planar. However, at equating the
colors of the gluons and the colors of initial and final quarks in the
baryon's lines, the color indexes in these diagrams flow as in
planar diagrams. Thereby their $N_{c}^{-1}\,$-behavior becomes as in
planar diagrams, as~well. Nevertheless, the diagrams fig.
\ref{Fig1}(e,f) are strongly suppressed because they imply double parton
scattering. We estimate their relative contribution to the cross-section
as 5\% of the contribution of diagrams with single parton scattering, by
analogy with the relative contribution of double parton scattering in
high-energy $pp$ and $p \bar p$ collisions \cite{DPP1,DPP2,DPP3}. So, in
the leading approximation, we are coming to consideration of only
processes of the types E and A.

In the case of sea-quark scattering, the relevant diagrams can be
obtained through splitting of the valence-quark lines in
fig.~\ref{Fig1}. The general rule is as follows: at the transition due
to the scattering of initial-state quarks (antiquarks) into the
final-state quarks (antiquarks) their charge should be increased per
unit in the composition of mesons, and decreased per unit in the
composition of baryons. With the $u$-, $d$-, $s$- quarks and antiquarks,
in the case of pure sea-quark scattering we thus  have for each of the
reactions in fours diagrams of the types E and A. In the case of
scattering of sea quarks on valence quarks we have along one diagram of
each type. Finally, in the case of scattering of valence quarks on sea
quarks for each of the reactions we have in twos diagrams of each type.
So, in total we have $(4+1+2) \times 2 \times 2 = 28$ diagrams. We do
not show them in view of triviality of the issue. What is important for
further analysis, is that for each reaction there are equal number of
diagrams of the types E and A.

The contributions of particular hard subprocesses can be easily
estimated. In doing so, we have to equate colors of quarks involved in
the scattering separately in the mesons and in the baryons in order to
compensate colors of spectators, and we have to equate helicities of
appropriate quarks in order to compensate helicities. Given these
conditions, direct calculation yields:  
\begin{equation}\label{text1}
M_{E} = N \times \hat{s}/\hat{u} \,,
 \qquad
M_{A} = N \times \hat{u}/\hat{s}\,.
\end{equation}
Here $M_{E}$ and $M_{A}$ are the amplitudes for diagrams with quark
exchange and with quark annihilation, $N$ is a common constant
proportional to the QCD coupling constant, $\hat{s}$ and $\hat{u}$ are
Mandelstam variables for subprocesses.

Based on formula (\ref{text1}), we conclude that in the limit of zero
transfer the amplitudes $M_{E}$ and $M_{A}$ coincide each other, and
are independent of the quarks flavors and energies. So the amplitudes
are common for the valence and sea quarks. At non-zero transfer the
amplitudes vary, but far weaker as compared to the amplitudes of real
particles. Actually their dependence on the transfer is covered by the
errors in spin-flip factors in the physical amplitudes. For this reason
it makes little sense to take into account the variations in $M_{E}$ and
$M_{A}$, at least in the range of transfer considered below. Further we
refer the mentioned variations to the account of soft-interaction
factors. 

In conclusion, we discuss how the above picture of scattering varies at
significantly increasing $t$. In this case the role of diagrams
fig. \ref{Fig1}(a-d) should fall since it becomes more and more
difficult for partons-spectators to turn round after the fast parton
that scatters on large angle. Simultaneously, the role of subprocesses
with double scattering becomes more and more important. We mean
subprocesses, when, for example, an antiquark in the incident meson
involves into the scattering at once two quarks in the proton, so that
as though a replacement occurs in the target of the proton by a neutral
meson. Such subprocesses become dominant in the case of backward
charge-exchange scattering. Below we do not consider this kinematic
region, and we mention only that the backward charge-exchange scattering
can be understood from the standpoint of the parton model, as well.

\section{Determination of the cross-sections}\label{sec3}

\begin{table}[t]
\caption{Charge-exchange reactions in the case of $\rho$--$a_2$
trajectories and $SU_{\!f}(3)$ coefficients in the vertex factors.
Right  column indicates underlying hard subprocesses.}\label{T1}
\begin{center}
\begin{tabular}{ c | c c | c c | c c  }
\hline\noalign{\medskip}
 $\qquad\qquad\qquad$    & \ $V$ \  \  & $SU(3)$
                         & \ $T$   & $\quad SU(3) \quad$    &
 $\quad$ hard     & subprocesses $\quad$
\\[1mm]
\hline\noalign{\medskip}
 $\pi^- p \to \pi^0 n$   & $\rho$      & $1$
                         & ---         & ---                &
 $\quad$ E $(u\bar u)$ & $\quad$ A $(d\bar d)$
\\[1mm]
\hline\noalign{\medskip}
 $\pi^- p \to \eta^8 n $ & ---         & ---
                         & $a_2$       & $ 1/\sqrt{3}$      &
 $\quad$ E $(u\bar u)$   & $\quad$ A $(d\bar d)$
\\[1mm]
\hline\noalign{\medskip}
 $\pi^- p \to \eta^0 n $ & ---         & ---
                         & $a_2$       & $\xi \sqrt{2/3}$   &
 $\quad$ E $(u\bar u)$   & $\quad$ A $(d\bar d)$
\\[1mm]
\hline\noalign{\medskip}
 $ K^- p \to \bar{K}^0 n $     & $\rho$      & $-1/\sqrt{2}$
                         & $a_2$       & $ 1/\sqrt{2}$   &
 $\qquad$                & $\quad$ A $(s\bar d)$
\\[1mm]
\noalign{\smallskip}\hline
\end{tabular}
\end{center}
\end{table}

\begin{table}[t]
\caption{The same that in Table \ref{T1} in the case of $K^*$--$K_2^*$
trajectories. }\label{T2}
\begin{center}
\begin{tabular}{ c | c c | c c | c c  }
\hline\noalign{\medskip}
 $\qquad\qquad\qquad$    & \ $V$ \  \  & $SU(3)$
                         & \ $T$   & $\quad SU(3) $    &
 $\quad$ hard     & subprocesses $\quad$
\\[1mm]
\hline\noalign{\medskip}
 $K^- p \to \pi^0 \Lambda$  & $K^{*}$     & $1/2$
                            & $K_2^{*}$   & $1/2$              &
 $\quad$ E $(u\bar u)$ & $\quad$
\\[1mm]
\hline\noalign{\medskip}
 $K^- p \to \eta^8 \Lambda$ & $K^{*}$     & $\sqrt{3}/2$
                            & $K_2^{*}$   & $\,-1/(2\sqrt{3})$ &
 $\quad$ E $(u\bar u)$ & $\quad$ A $(s\bar s)$
\\[1mm]
\hline\noalign{\medskip}
 $K^- p \to \eta^0 \Lambda$ & ---         & ---
                            & $K_2^{*}$   & $\xi \sqrt{2/3}$   &
 $\quad$ E $(u\bar u)$ & $\quad$ A $(s\bar s)$
\\[1mm]
\hline\noalign{\medskip}
 $\pi^- p \to K^0 \Lambda$  & $K^{*}$    & $-1/\sqrt{2}$
                         & $K_2^{*}$     & $ 1/\sqrt{2}$   &
 $\quad$                & $\quad$ A $(d\bar s)$
\\[1mm]
\noalign{\smallskip}\hline
\end{tabular}
\end{center}
\end{table}

A modern approach for the description of charge-exchange reactions
involves the use of Regge phenomenology, which in turn is based on an
idea of exchanges by appropriate trajectories in the $t$-channel
\cite{Regge,SU3}. In the case of $\pi^- p \to M^0 n$ and $K^- p \to M^0
\Lambda$ at not too large $|t|$, the leading are the $\rho$, $a_2$ and
$K^*$, $K_2^*$ trajectories, see table~\ref{T1} and table~\ref{T2},
respectively.\footnote{With superfluously increasing $|t|$ the
contributions of Regge cuts become dominant. On the opposite edge of the
kinematic region, where $|u|$ is sufficiently small, the $N$-trajectory
dominates.} The tables include also other charge-exchange reactions with
the same trajectories, and the $SU_{\!f}(3)$ coefficients that emerge in
the vertex factors in the assumption that reactions are mediated by
octet exchanges. The gaps in the tables mean absence of contributions.
Parameter $\xi$ describes possible violation of the quark symmetry in
isosinglet channel. As discussed~above, the charge-exchange reactions
occur through the hard-scattering subprocesses. An information about
them is given in the last columns in the tables. Recall that by E and
A we indicate different types of hard subprocesses. In brackets we give
their alternative identification on the basis of valence quarks in the
final state, {\it cf.}~fig.~{\ref{Fig1}(a-d). 

Our aim now is to link hard and soft contributions. We do this based on
the property of independence of hard-scattering contributions from the
flavors and types (valence, sea) of relevant quarks. Their relative
weights are inessential, as well, as they are independent of energy.
Additionally we assume that soft contributions are basically independent
of flavors and types of hard-scattered quarks. However, soft
contributions are sensitive to the flavors of the valence quarks in the
final state and to the channel in which the hard virtuality is
transferred. Especially they may be sensitive to the sign of the hard
virtuality, which is negative and positive in the cases of E- and A-type
contributions, respectively. On this basis, we consider joint hard and
soft contributions in the dependence on types of underlying hard
subprocesses or, which is equivalent, in the dependence on flavors of
the valence quarks in the final state.

On this basis we further consider separately the cases of coherent and
incoherent summation of intermediate contributions. In the case of
coherent summation the hard contributions appear in the amplitude in
the form of superpositions. In relation to the valence-quark content in
the final state they may be symmetric or antisymmetric. Analysis of the
tables shows that the relative sign of individual contributions obeys
the following rule: the E-type contributions appear always with positive
sign, while the A-type contributions are positive with the tensor
exchanges and negative with the vector exchanges (more precisely, with
the even or odd angular momentum of the $t$-channel exchanges). This
means that with the tensor exchanges the hard contributions appear in
symmetric form, while with the vector exchanges they appear in
antisymmetric form. Formally this may be written in the form
\begin{eqnarray}\label{text2}
 \frac{1}{\sqrt{2}} (u \bar u - d \bar d)_V &=&
 \pi^0, 
\\ \label{text3}
 \frac{1}{\sqrt{2}} (u \bar u + d \bar d)_T &=&
   \sqrt{\frac{2}{3}}\;\eta^0 + \sqrt{\frac{1}{3}}\;\eta^8\,,
\\ \label{text4}
 \frac{1}{\sqrt{2}} (u \bar u - s \bar s)_V &=&
 \frac{1}{2}\;\pi^0   + \frac{\sqrt{3}}{2}\;\eta^8,
\\ \label{text5}
 \frac{1}{\sqrt{2}} (u \bar u + s \bar s)_T &=&
 \frac{1}{2}\;\pi^0   + \sqrt{\frac{2}{3}}\;\eta^0 -
 \frac{1}{2\sqrt{3}}\;\eta^8\,. \quad
\end{eqnarray}
Here $1\!/\!\sqrt{2}$ in the l.h.s.~is a common factor, indices V and T
indicate the trajectory in the presence of which given contributions
appear in the amplitude. In the r.h.s.~the same expressions are
represented in the standard $SU_{\!f}(3)$ basis. What is important in
relations (\ref{text2})--(\ref{text5}) is that the weights in the
r.h.s.~reproduce the group coefficients in the tables. 

However we have omitted parameter $\xi$ in the above discussion.
Generally its origin can be connected with both the soft and hard
interactions. In the former case it means a difference of the couplings
of trajectories with octet and pure singlet states. Due to hard
interactions, the violation of isosinglet symmetry can arise from
diagrams fig.~\ref{Fig1}(e,f) taken with the option of transition of
``outgoing'' gluons into $\eta^0$. In this case $\xi$ is constrained in
the assumption that appropriate contributions do not exceed 5\% in the
cross-section (see sect.~\ref{sec2}). In amplitude this means 2.5\%,
which gives $\xi = 1 \pm 0.025$.

We have discussed above the mode of coherent summation of hard
contributions. In the opposed case when the coherence is completely
lost, the hard contributions must be summed up in the cross-section. In
this case the $SU_{\!f}(3)$ coefficients in the vertex factors are
determined by the relations, 
\begin{eqnarray}\label{text6}
 \frac{1}{\sqrt{2}} \; (u \bar u)
 &=&
 \frac{1}{\sqrt{6}}\;\eta^0 + \frac{1}{\sqrt{12}}\;\eta^8 +
 \frac{1}{2}\;\pi^0 \,,
\\[0.5\baselineskip] \label{text7}
 \frac{1}{\sqrt{2}} \; (d \bar d)
 &=&
 \frac{1}{\sqrt{6}}\;\eta^0 + \frac{1}{\sqrt{12}}\;\eta^8 -
 \frac{1}{2}\;\pi^0 \,,
\\[0.5\baselineskip] \label{text8}
 \frac{1}{\sqrt{2}} \; (s \bar s)
 &=&
 \frac{1}{\sqrt{6}}\;\eta^0 - \frac{1}{\sqrt{3}}\;\eta^8 \,.
\end{eqnarray}
In so doing one should remember that in some channels the contributions
are forbidden by the quantum numbers of the initial and final states
\cite{Regge}, see the gaps in the tables.

Now we can define the cross-sections of real processes. Formally we do
this by using Regge parameterization, but simultaneously we take into
consideration the above results about the mode of summation, and the
$SU_{\!f}(3)$ coefficients in the vertex factors. Recall that the vertex
factors are responsible for the couplings of trajectories with the
``in'' and ``out'' states. It is typically assumed that within each
trajectory they are common at zero transfer (up to $SU_{\!f}(3)$
coefficients). However, with growing $|t|$ the vertex factors may
individually vary depending on specific initial and final states. Next
the key point is the presence in the amplitudes of signature factors.
They include multiplier $\mbox{i}\times\sin(\pi \alpha_V /2)$ in the
case of V-exchanges relative to multiplier $\cos(\pi \alpha_T /2)$ in
the case of T-exchanges, where $\alpha_V=\alpha_V(t)$ and
$\alpha_T=\alpha_T(t)$ are the trajectories. Notice that since the
contributions of V- and T-exchanges are mutually imaginary, they do not
interfere with each other. 

The above information is sufficient for constructing the cross-sections.
What additionally is required is a mixing of isosinglet states. In what
follows unless otherwise specified, we use a simplest
model for the $\eta$--$\eta'$ mixing \cite{PDG}: 
\begin{eqnarray}\label{text9}
\! |\,\eta \,>\, &=&
 \cos\theta \; |\,\eta^8> - \sin\theta \; |\,\eta^0>\,,
 \nonumber\\ [-0.5\baselineskip]
 && \\ [-\baselineskip]
 \nonumber\\ [-0.5\baselineskip]
 |\,\eta'> &=&
 \sin\theta \; \,|\,\eta^8> + \cos\theta \; |\,\eta^0>\,.
 \nonumber
\end{eqnarray}
Here $|\,\eta^8>$ and $|\,\eta^0>$ mean states that match the
$\eta^8$ and $\eta^0$ in (\ref{text2})--(\ref{text8}).

\subsection{\mbox{\boldmath$\rho$}--\mbox{\boldmath$a_2$}
trajectories}\label{sec3.1}

On the basis of table~\ref{T1} and formulas (\ref{text2}), (\ref{text3})
(\ref{text6}), (\ref{text7}), (\ref{text9}) we immediately obtain the
following expressions for the differential cross-sections for $\pi^{-} p
\to M^{0}
n$, $M^0 =\pi^0, \eta, \eta'$,  
\begin{eqnarray}\label{text10}
 \sigma (\pi^{-} p \to \pi^{0} n) &=& \gamma \, g_{\pi\rho\pi^0}^2 \,
 \sin^2 \!\frac{\pi \alpha_{\rho}}{2}
 \left(\frac{s}{s_0}\right)^{2\alpha_{\rho}-2} ,
\\ [0.2\baselineskip] \label{text11}
 \sigma (\pi^{-} p \to \eta  n)\, &=&
 \frac{\gamma}{3} \; g_{\pi a_2 \eta }^2 \,
 \cos^2 \!\frac{\pi \alpha_{a_2}}{2}
 \left(\cos\theta - \xi\sqrt{2}\sin\theta\right)^2
 \left(\frac{s}{s_0}\right)^{2\alpha_{a_2}-2}  ,
\\ [0.2\baselineskip] \label{text12}
 \sigma (\pi^{-} p \to \eta' n) &=&
 \frac{\gamma}{3} \; g_{\pi a_2 \eta'}^2 \,
 \cos^2 \!\frac{\pi \alpha_{a_2}}{2}
 \left(\sin\theta + \xi\sqrt{2}\cos\theta\right)^2
 \left(\frac{s}{s_0}\right)^{2\alpha_{a_2}-2}  .
\end{eqnarray}
Hereinafter we mean  ${\rm d}\sigma \!/\! {\rm d}t$ by the $\sigma$.
In the above formulas $g_{\pi\rho\pi}$ and $g_{\pi a_2 \eta(\eta')}$
are the vertex factors (without $SU_{\!f}(3)$ coefficients), $s_0$ is a
dimensional parameter. Parameter $\gamma$ takes values 1 or 1/2 in the
cases of coherent or incoherent summation of hard contributions,
respectively. For simplicity we consider spin-flip factors absorbed by
the vertex factors.

Trivial manipulations lead to a more convenient form for (\ref{text11})
and (\ref{text12}):
\begin{eqnarray}\label{text13}
 \sigma (\pi^- p \to \eta  n) &=&
 \frac{1\!+\!2\xi^2}{3}\;\gamma \, g_{\pi a_2 \eta }^2 \,
 \cos^2 \!\frac{\pi \alpha_{a_2}}{2} \,
 \cos^2(\theta + \theta_{id} - \delta)
 \left(\frac{s}{s_0}\right)^{2\alpha_{a_2}-2} ,
\\ [0.2\baselineskip] \label{text14}
 \sigma (\pi^- p \to \eta' n) &=&
 \frac{1\!+\!2\xi^2}{3}\;\gamma \, g_{\pi a_2 \eta'}^2 \,
 \cos^2 \!\frac{\pi \alpha_{a_2}}{2} \,
 \sin^2(\theta + \theta_{id} - \delta)
 \left(\frac{s}{s_0}\right)^{2\alpha_{a_2}-2} .
\end{eqnarray}
Here $\theta_{id}$ is the so called ideal mixing angle,
$\theta_{id} = \arctan\!\sqrt{2}$ \ ($\theta_{id} \approx
54.7^0$), and
\begin{equation}\label{text15}
\delta = \arctan \frac{\sqrt{2}(1-\xi)}{1+2\xi}\,.
\end{equation}
With $\xi = 1 \pm 0.025$, we have $|\delta| < 0.7^0$.

Assuming that the coupling of $a_2$-trajectory with isosinglet mesons is
common at zero transfer, we get the ratio of the cross-sections, 
\begin{equation}\label{text16}
 R_{\pi}^{\eta'/\eta}(0) \; \equiv \; \left.
 \frac{\sigma(\pi^- p \to \eta' n)}
 {\sigma(\pi^- p \to \eta  n)}\right\vert_{ \;t=0}  = \;
 \tan^2 (\theta + \theta_{id} - \delta)\,.
\end{equation}
This ratio was measured in different experiments. The more precise value
was obtained by GAMS-4$\pi$ \cite{GAMS}, with
\begin{equation}\label{text17}
R_{\pi}^{\eta'/\eta}(0) = 0.54 \pm 0.04 \,.
\end{equation}
This result may be compared with that of NICE, $R_{\pi}^{\eta'/\eta}(0)
= 0.55 \pm 0.06$ \cite{NICE1}, and that of Argonne ZGS,
$R_{\pi}^{\eta'/\eta}(0) = 0.500$ $\pm 0.092$ \cite{Stanton}. On the
basis of (\ref{text16}) and (\ref{text17}) we get    
\begin{equation}\label{text18}
 \theta \;=\; (- 18.4^0  \pm 1.0)^0 .
\end{equation}
Here we show experimental error only. 

Our result (\ref{text18}) nearly coincides with corresponding result of
\cite{NICE1}. However, it sharply differs from that of \cite{Stanton},
in spite of compatible $R_{\pi}^{\eta'/\eta}(0)$. The reason is that
\cite{Stanton} used another ratio for the definition of $\theta$, namely
(as declared in \cite{Stanton}) the ratio of spin-flip contributions to
the cross-sections at zero transfer. However, spin-flip contributions
vanish in this limit, and therefore their ratio cannot be directly
measured at $t=0$. So what actually was used in \cite{Stanton} for the
definition of $\theta$, was the ratio of certain integrals of the
differential cross-sections over a finite range of $t$. Generally this
means a loss of connection with the true flavor content of $\eta$ and
$\eta'$. For this reason we believe that approach \cite{Stanton} for
definition of $\theta$ is incorrect. 

\begin{figure}[t]
\hbox{ \hspace*{95pt}
       \epsfxsize=0.4\textwidth \epsfbox{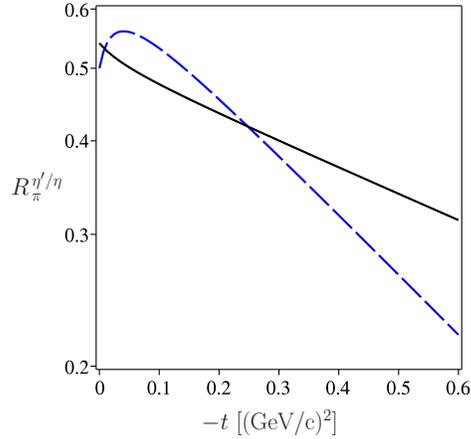}}
\caption{\small $R_{\pi}^{\eta'/\eta}(t)$ defined on the basis of
data \cite{GAMS} (solid curve) and \cite{Stanton} (dashed curve).}
\label{Fig2}
\end{figure}

At nonzero transfer, $R_{\pi}^{\eta'/\eta}(t)$ has
nontrivial behavior,
which is determined by the behavior of the vertex factors, 
\begin{equation}\label{text19}
 R_{\pi}^{\eta'/\eta}(t) \; = \;R_{\pi}^{\eta'/\eta}(0)
 \left[
 \frac{g_{\pi a_2 \eta'}(t)}{g_{\pi a_2 \eta }(t)}
 \right]^2 \,.
\end{equation}
As an illustration, in Fig.~\ref{Fig2} we show $R_{\pi}^{\eta'/\eta}(t)$
measured in \cite{GAMS} and \cite{Stanton}. At not too small $|t|$,
where contributions of spin-flip factors become irrelevant, in both
cases $R_{\pi}^{\eta'/\eta}(t) \sim \exp(c_R t)$, but the slopes $c_R$
are different. Namely $c_R=1.87 \pm 0.22$ (GeV$\!$/c)$^{-2}$ in
\cite{Stanton} and $c_R=0.80 \pm 0.22$ (GeV$\!$/c)$^{-2}$ in
\cite{GAMS}. Their difference is $1.07 \pm 0.31$ (GeV$\!$/c)$^{-2}$,
which differs from zero by more than 3$\sigma$. This means that soft
contributions undergo changes at the transition to higher energies (from
the beam momentum 8.45 GeV$\!$/c in \cite{Stanton} to 32.5 GeV$\!$/c in
\cite{GAMS}).

Finally, assuming that at $t=0$ the vertex factors are common within the
$\rho$- and within the $a_2$-trajectories, we can extract their ratio.
Notice, this issue is of independent interest since it allows us to
check whether the $\rho$- and $a_2$-trajectories are degenerate. So
putting $g_{\pi \rho \pi^0}(0) = g_{\rho 0}$, $g_{\pi a_2 \eta (')}(0) =
g_{a_2 0}$, on the basis of (\ref{text10})--(\ref{text12}) we obtain
\begin{equation}\label{text20}
 \left. \frac{\sigma (\pi^{-} p \to \pi^{0} n)}
 {\sigma (\pi^{-} p \to \eta  n) + \sigma (\pi^{-} p \to \eta' n)}
 \right\vert_{ \;t=0}
 = \;  \frac{3}{1+2\xi^2}
 \left[\frac{g_{\rho 0}\,\sin\left[\pi \alpha_{\rho 0}/2\right]}
            {g_{a_2 0}\,\cos\left[\pi \alpha_{a_2 0}/2\right]} \right]^2
 \left(\frac{s^2}{s_0^2}\right)^{\alpha_{\rho 0}-\alpha_{a_2 0}} .
\end{equation}
The intercepts of the trajectories are known, $\alpha_{\rho 0} = 0.48
\pm 0.01$, $\alpha_{a_2 0} = 0.38 \pm 0.02$ \cite{NICE2,NICE3}. So the
ratio of the couplings can be extracted. To our regret, there are
no data at low energies, and with sufficiently small errors the data are
only available at $\pi^-$ momenta above 15 GeV$\!$/c
\cite{NICE1,NICE2,NICE3}. From this we obtain $g_{\rho 0}/g_{a_2 0} =
4.0 \pm 1.1$.

\subsection{\mbox{\boldmath $K^*$}--\mbox{\boldmath $K^*_2$}
trajectories}\label{sec3.2}

Unfortunately, the $K^*$- and $K^*_2$-trajectories are not quite
accurately measured. For this reason we put $\alpha_{K^{*}_2} =
\alpha_{K^{*}}$, and we use a calculated trajectory $\alpha_{K^{*}}(t) =
0.32 + 0.84 \, t$ \cite{Ebert}. In the case of the coherent summation of
hard contributions on the basis of table~\ref{T2} and formulas
(\ref{text4}), (\ref{text5}), (\ref{text9}), we have 

\begin{eqnarray}\label{text21}
 \sigma (K^{-} p \to \pi^{0}\Lambda) &=&
 \frac{1}{4} \left[
 g_{K K^{*}  \pi^0}^2 \,
 \sin^2 \!\frac{\pi \alpha_{K^{*}}}{2} \;+\;
 g_{K K^{*}_2\pi^0}^2 \,
 \cos^2 \!\frac{\pi \alpha_{K^{*}}}{2} \right]
 \left(\frac{s}{s_0}\right)^{2\alpha_{K^{*}}-2} \!\!,\qquad
\end{eqnarray}
\begin{eqnarray}\label{text22}
 \sigma (K^{-} p \to \eta  \Lambda) &=&
 \frac{3}{4} \left[
 g_{K K^{*}  \eta }^2 \, \sin^2 \!\frac{\pi \alpha_{K^{*}}}{2}
 \cos^2 \theta  \right. \;+\;
 \frac{1\!+8\xi^2}{9} \times
 \nonumber\\
&&
 \left. \quad\;
 g_{K K^{*}_2\eta }^2 \, \cos^2 \!\frac{\pi \alpha_{K^{*}}}{2}
 \cos^2(\theta + \theta'_{id} - \delta') \right]
 \left(\frac{s}{s_0}\right)^{2\alpha_{K^{*}}-2} \!\!\!,\qquad
\\ [0.7\baselineskip] \label{text23}
 \sigma (K^{-} p \to \eta' \Lambda) &=&
 \frac{3}{4} \left[
 g_{K K^{*}  \eta'}^2 \, \sin^2 \!\frac{\pi \alpha_{K^{*}}}{2}
 \sin^2 \theta \right. \;+\;
 \frac{1\!+8\xi^2}{9} \times
 \nonumber\\
&&
 \left. \quad \;
 g_{K K^{*}_2\eta'}^2 \, \cos^2 \!\frac{\pi \alpha_{K^{*}}}{2}
 \sin^2(\theta + \theta'_{id} - \delta') \right]
 \left(\frac{s}{s_0}\right)^{2\alpha_{K^{*}}-2} \!\!\!.\qquad
\end{eqnarray}
Here $\theta'_{id} = -\arctan (2\sqrt{2})$, $\theta'_{id} \approx
-70.5^0$, and
\begin{equation}\label{text24}
\delta' = -\arctan \frac{2\sqrt{2}(1-\xi)}{1+8\xi}\,.
\end{equation}
With $\xi = 1\!\pm\!0.025$, $(1\!+8\xi^2)/9 = 1\!\pm\!0.045$ and
$|\delta'|  <  0.5^0$.

The presence of two trajectories does not allow us to extract
independently the mixing angle. However, knowing $\theta$, we can
extract the ratio of the couplings of $K^*$- and $K^*_2$-trajectories
with pseudoscalar mesons. Putting $g_{K K^{*}\eta(')}(0) = g_{K^{*} 0}$
and $g_{K K^{*}_2\eta(')}(0) = g_{K^{*}_2 0}$, from (\ref{text22}) and
(\ref{text23}) we get
\begin{equation}\label{text25}
 R_{K}^{\eta'/\eta}(0) \; \equiv \; \left.
 \frac{\sigma(K^{-} p \to \eta' \Lambda)}
 {\sigma(K^{-} p \to \eta \Lambda)}\right\vert_{ \;t=0}  = \;
 \frac{
       r_0 \, \tan^2\!\frac{\pi \alpha_{K^{*}0}}{2} \cos^2 \theta +
              \cos^2(\theta + \theta'_{id} - \delta')}
      {
       r_0 \, \tan^2\!\frac{\pi \alpha_{K^{*}0}}{2} \sin^2 \theta +
              \sin^2(\theta + \theta'_{id} - \delta')} \,.
\end{equation}
Here $r_0 = g_{K^{*} 0}^2/g_{K^{*}_2 0}^2 \times 9/(1\!+8\xi^2)$. At
relatively low energies the ratio (\ref{text25}) was measured at 8.25
GeV$\!$/c  \cite{Harran},
\begin{equation}\label{text26}
 R_{K}^{\eta'/\eta}(0) = 1.37 \pm 0.13 \,.
\end{equation}
Taking into account (\ref{text18}), (\ref{text25}), and (\ref{text26}),
we obtain
\begin{equation}\label{text27}
 g_{K^{*} 0}/g_{K^{*}_2 0} = 1.71 \pm 0.12 \,.
\end{equation}

At nonzero $t$, formulas (\ref{text21})--(\ref{text23}) describe well
appropriate data under assumption that in each reaction the~ver\-tex
factors have common exponential behavior. In fig.~\ref{Fig3}(a) and
fig.~\ref{Fig3}(b) we show data for $K^{-} p \to \pi^{0}\Lambda$ at 4.2
GeV$\!$/c \cite{Marzano} and 8.25 GeV$\!$/c \cite{Harran}, respectively.
In the same places we plot theoretical curve (\ref{text21}) with the
vertex factors $g^2(t) \sim \exp(ct)$ where $c$ is varying with the
energy in accordance with Regge parameterization. Namely, we put $c_a -
c_b = 2\,\alpha' \ln (s_a/s_b)$, where $a$ and $b$ mark different
energies and $\alpha'$ is the slope of the trajectory, $\alpha(t) =
\alpha_0 + \alpha' t$. 

\begin{figure}
 \includegraphics[width=0.4\textwidth]{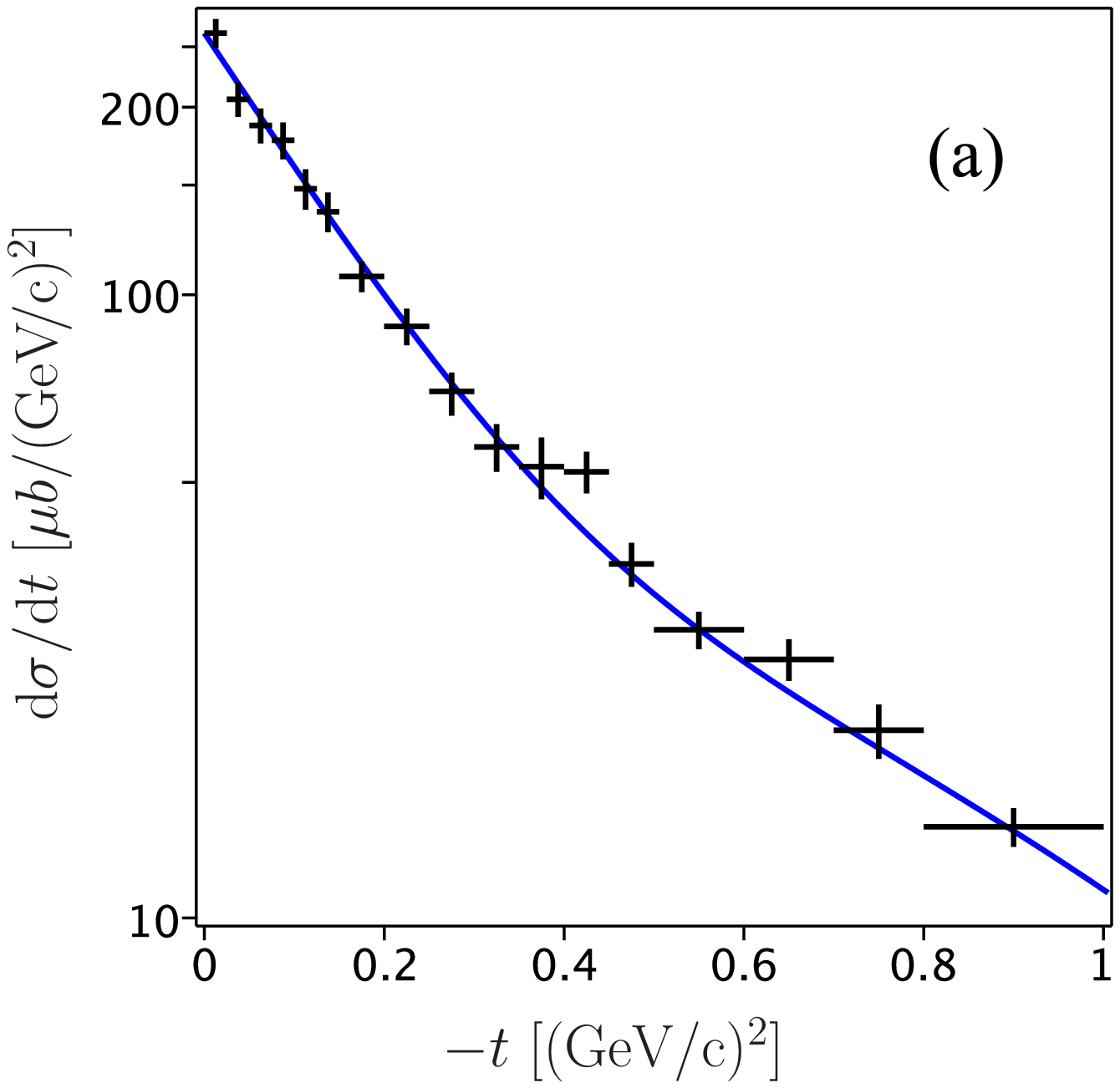}
 \hfill 
 \includegraphics[width=0.4\textwidth]{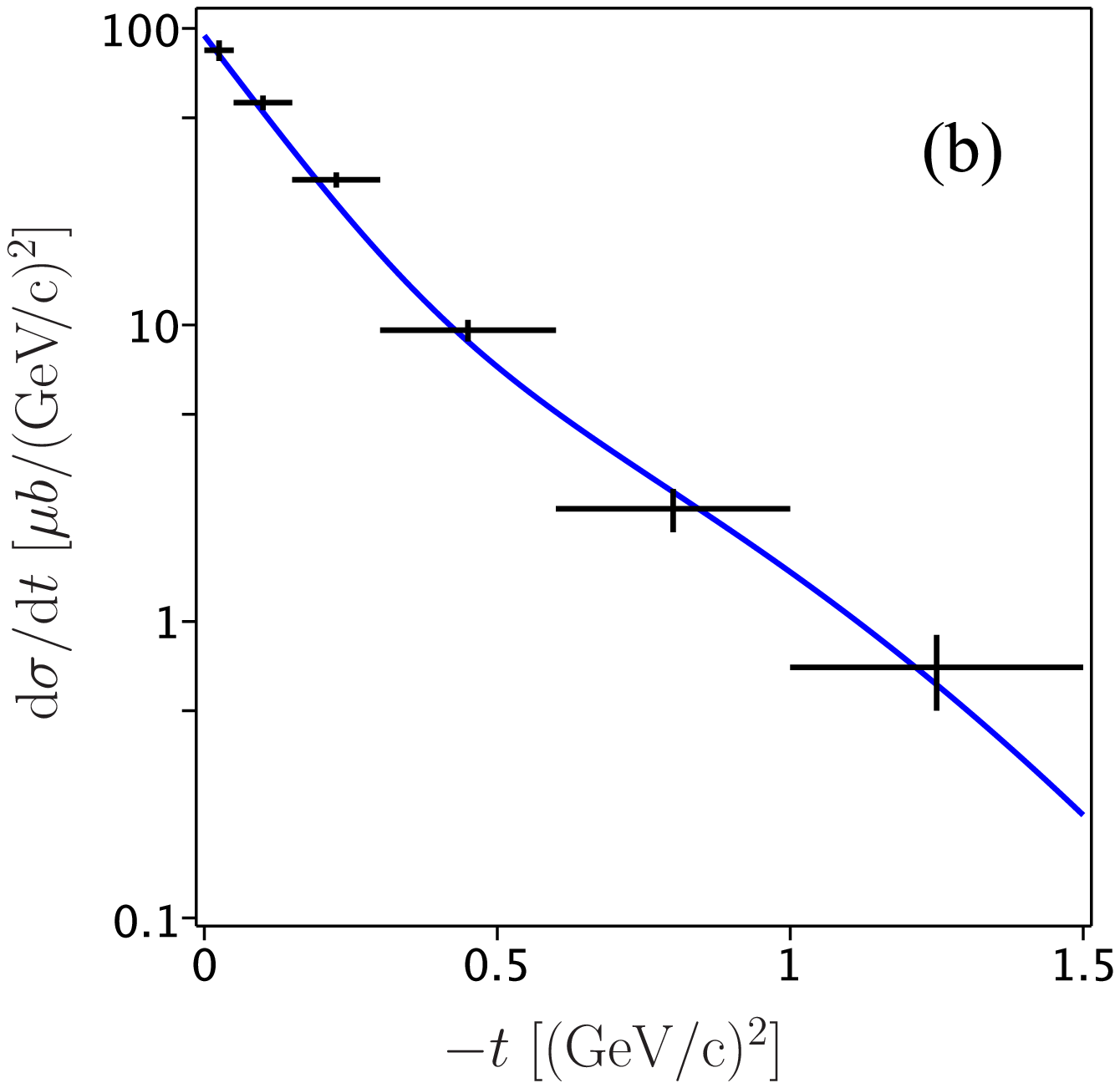}
\caption{\small Differential cross-section $K^{-} p \to \pi^{0}\Lambda$
at 4.2 GeV$\!$/c \cite{Marzano} (a), and 8.25 GeV$\!$/c \cite{Harran}
(b). Theoretical curves correspond to coherent summation of hard
contributions.
} \label{Fig3}
\end{figure}
\begin{figure}[p]
 \includegraphics[width=0.4\textwidth]{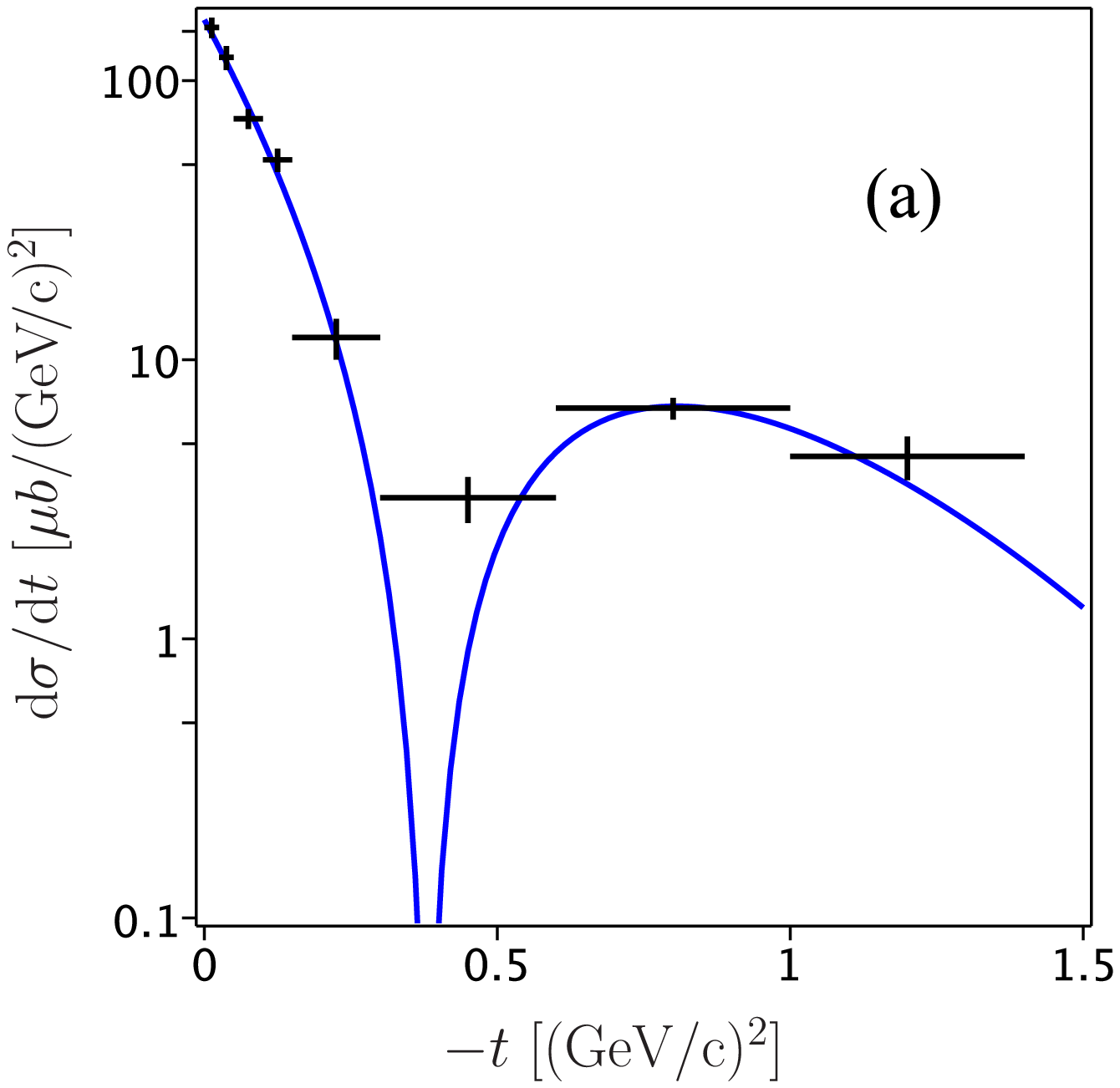}
 \hfill
 \includegraphics[width=0.4\textwidth]{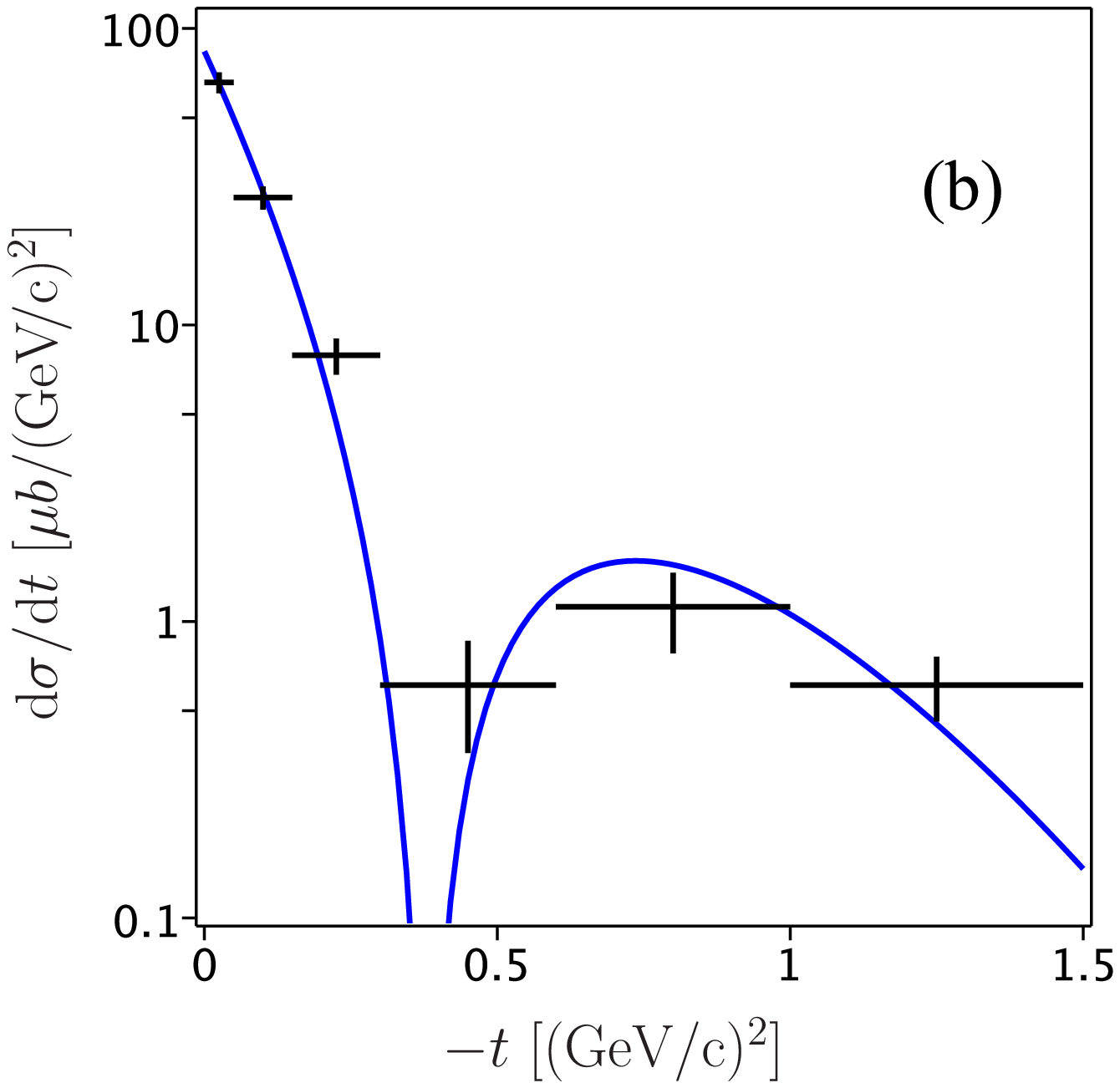}
\caption{\small The same that in Fig.\ref{Fig3} for $K^{-} p \to
\eta\Lambda$.
} \label{Fig4}
\end{figure}
\begin{figure}[p]
 \includegraphics[width=0.4\textwidth]{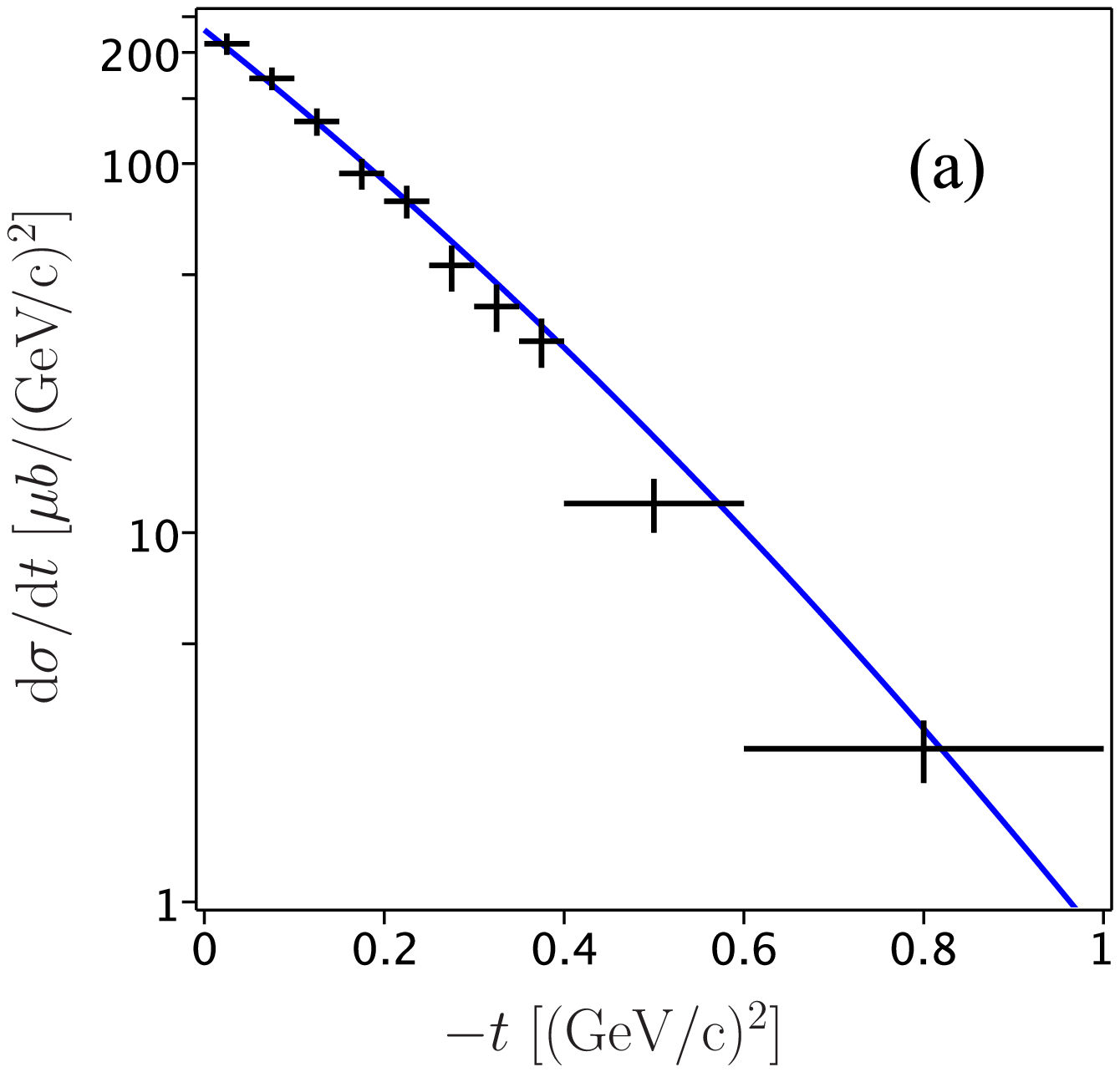}
 \hfill
 \includegraphics[width=0.4\textwidth]{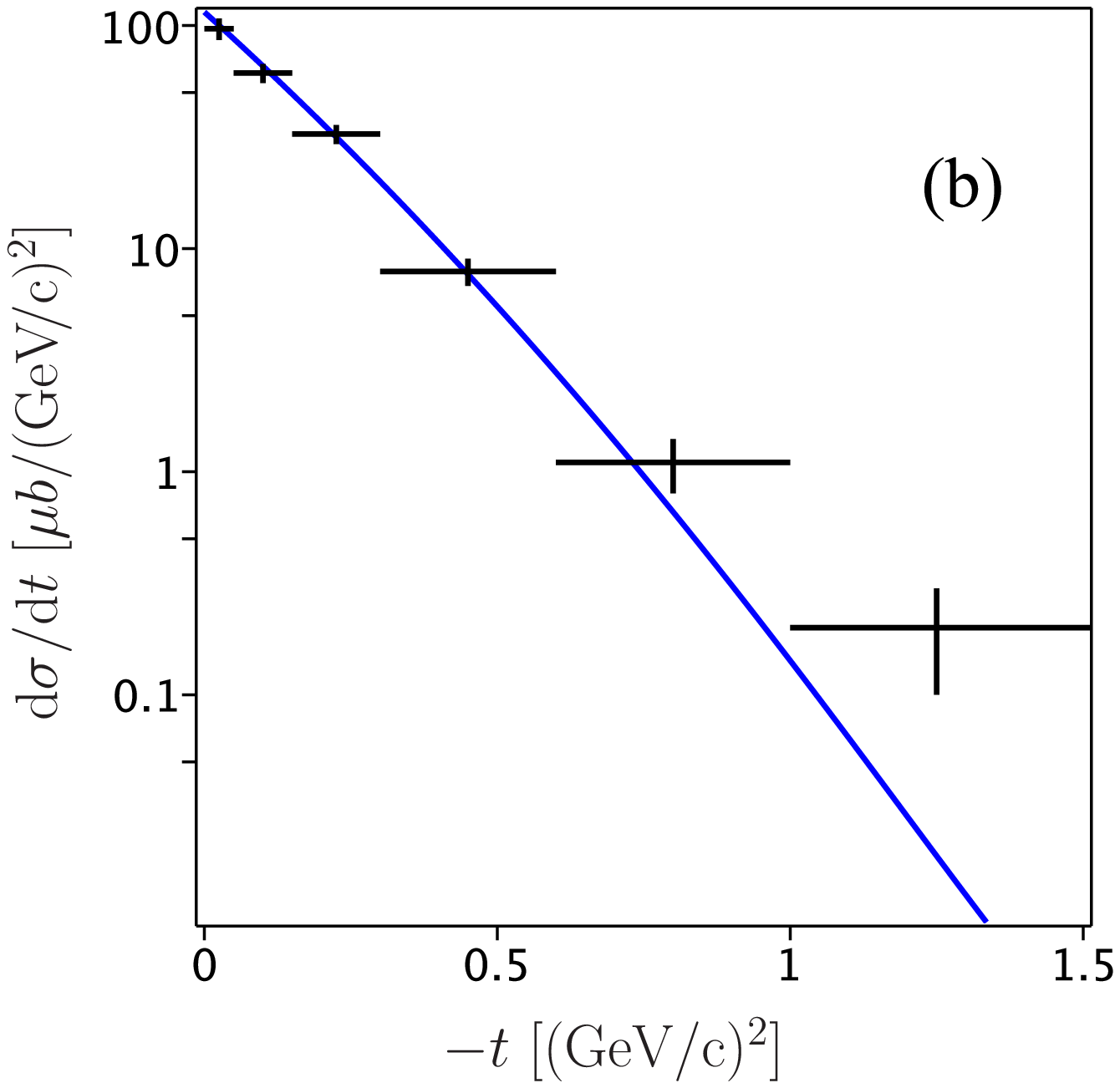}
\caption{\small The same that in Fig.\ref{Fig3} for $K^{-} p \to
\eta'\Lambda$.
} \label{Fig5}
\end{figure}

In fig.~\ref{Fig4}(a,b) and fig.~\ref{Fig5}(a,b) in a similar manner we
show data \cite{Marzano,Harran} and corresponding theoretical curves
in~the cases of $\eta$ and $\eta'$ production. A notable feature of the
data in fig.~\ref{Fig4} is a pronounced dip near $-t \approx$ 0.4--0.5
(GeV$\!$/c)$^2$. In our approach it arises from the zero of the $K^*$
exchange contribution when $\alpha_{K^{*}}(t)=0$, coupled with the
suppression of $K^*_2$ relative to $K^*$ exchange by a factor
$\cos^2(\theta + \theta'_{id} - \delta') \approx 0.0003$. Previously,
this effect was discovered in \cite{Martin} in the framework of the
exchange-degenerate pole model. Any other explanation for this effect
was not found. 

\begin{figure}
\hbox{ \hspace*{95pt}
       \epsfxsize=0.45\textwidth \epsfbox{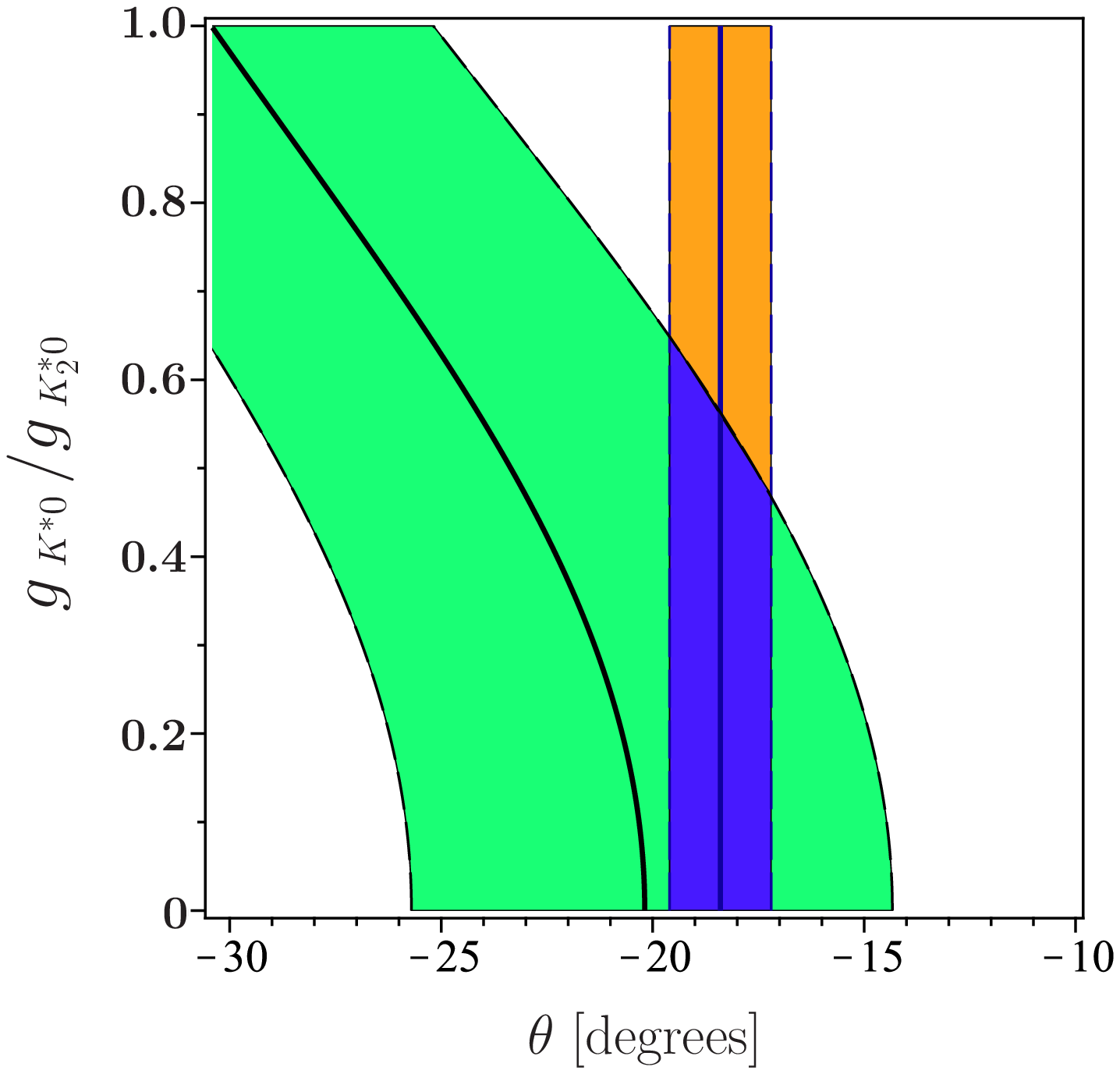}}
\caption{\small Allowed regions for $g_{K^{*} 0}/g_{K^{*}_2 0}$ and 
$\theta$.} \label{Fig6}
\end{figure}

In the case of incoherent summation of hard contributions, formula
(\ref{text21}) remains unchanged, but instead of (\ref{text22}) and
(\ref{text23}) in accordance with (\ref{text6}), (\ref{text8}),
(\ref{text9}) we have 
\begin{eqnarray}\label{text28}
& \displaystyle
 \sigma (K^{-} p \to \eta  \Lambda) =
 \left\{ \frac{5}{12} \, g_{K K^{*}  \eta }^2 \,
 \sin^2 \!\frac{\pi\alpha_{K^{*}}}{2} \, \cos^2 \theta +
         \frac{1}{4}  \, g_{K K^{*}_2\eta }^2 \,
 \cos^2 \!\frac{\pi \alpha_{K^{*}}}{2} \times
 \right. \qquad\qquad
&
 \nonumber\\ [0.2\baselineskip]
& \displaystyle 
  \biggl. \left[
      \frac{1\!+\!2\xi^2}{3}   \cos^2(\theta+\theta_{id}-\delta) +
 2\xi^2\frac{1\!+\!2\xi^{-2}}{3} \sin^2(\theta+\theta_{id}-\tilde\delta)
 \right]\biggr\}
  \left(\frac{s}{s_0}\right)^{2\alpha_{K^{*}}-2} \!\! , 
&
\\ [1.2\baselineskip] \label{text29}
& \displaystyle
 \sigma (K^{-} p \to \eta' \Lambda) =
 \left\{ \frac{5}{12}\, g_{K K^{*}  \eta'}^2 \,
 \sin^2 \!\frac{\pi\alpha_{K^{*}}}{2} \, \sin^2 \theta +
        \frac{1}{4}  \, g_{K K^{*}_2\eta'}^2 \,
 \cos^2 \!\frac{\pi \alpha_{K^{*}}}{2} \times
 \right. \qquad\qquad
&
 \nonumber\\ [0.2\baselineskip]
& \displaystyle 
  \biggl. \left[
      \frac{1\!+\!2\xi^2}{3}   \sin^2(\theta+\theta_{id}-\delta) +
 2\xi^2\frac{1\!+\!2\xi^{-2}}{3} \cos^2(\theta+\theta_{id}-\tilde\delta)
 \right]\biggr\}
  \left(\frac{s}{s_0}\right)^{2\alpha_{K^{*}}-2} \!\! . 
&
\end{eqnarray}
Here $\delta$ is defined in (\ref{text15}), and
\begin{equation}\label{text30}
\tilde\delta = \arctan \frac{\sqrt{2}\,(1-\xi^{-1})}{1+2\,\xi^{-1}}\,.
\end{equation}

From the ratio of the cross-sections, we again extract an information
about the ratio of the vertex factors. This time we use data at 32.5
GeV$\!$/c, which gives
\cite{GAMS} 
\begin{equation}\label{text31}
 R_{K}^{\eta'/\eta}(0) = 1.27 \pm 0.15 \,.
\end{equation}
Equating theoretical $R_{K}^{\eta'/\eta}(0)$ (we omit corresponding
formula) to the r.h.s. in (\ref{text31}), we get a corridor of allowed
values of $g_{K^{*} 0}/g_{K^{*}_2 0}$ vs $\theta$. We plot this in
fig.~\ref{Fig6}, where simultaneously we plot constraint (\ref{text18})
for $\theta$. The intersectional area shows allowed region for $g_{K^{*}
0}/g_{K^{*}_2 0}$. It is readily seen that it is bounded from above.
Approximately we have
\begin{equation}\label{text32}
 g_{K^{*} 0}/g_{K^{*}_2 0} \; \lesssim \; 0.6 \,.
\end{equation}
This result is to be compared with (\ref{text27}). The difference
can be treated as a consequence of the mode change of summation of
intermediate contributions.

\begin{figure}
 \includegraphics[width=0.4\textwidth]{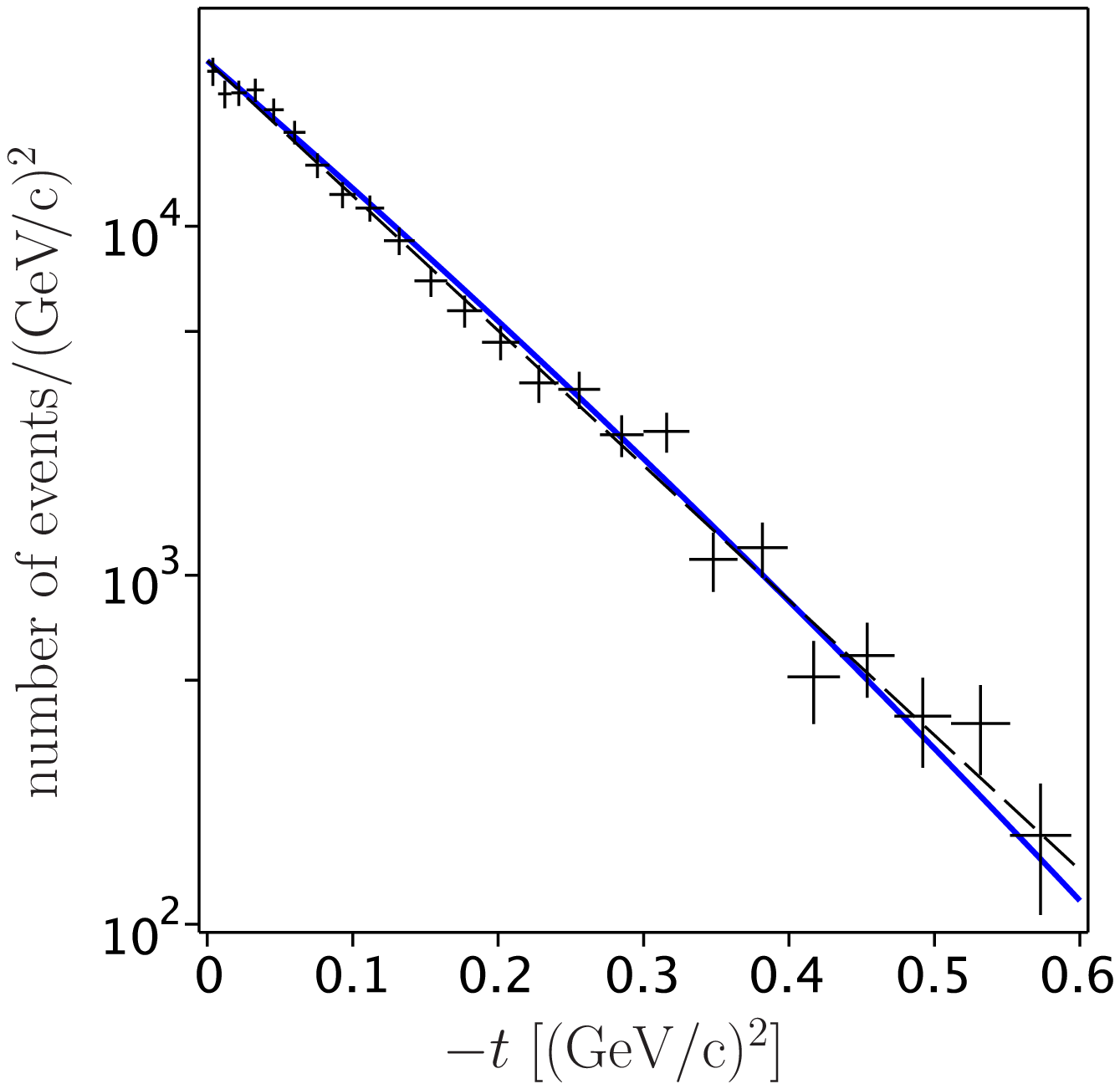}
 \hfill 
 \includegraphics[width=0.4\textwidth]{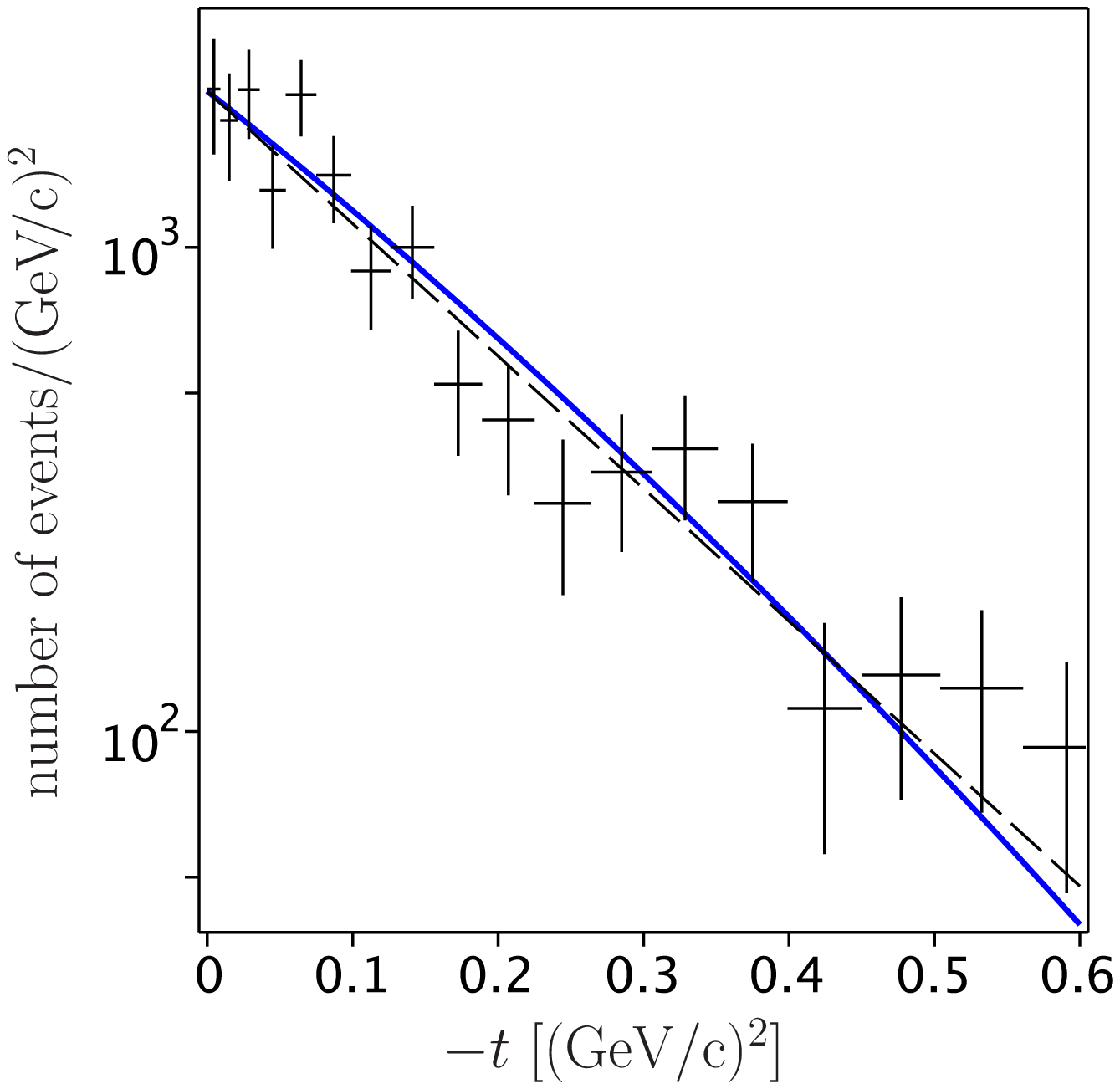}
\parbox[t]{0.47\textwidth}{
\vspace*{-.5\baselineskip}
 \caption{\small $\mbox{d}\!N/\mbox{d}t$ for $K^{-} p \!\!\to\!\!
\eta\Lambda$
at 32.5 GeV$\!$/c. Data and the exponential fit (dashed line) are taken
in \cite{GAMS}. The theoretical curve (solid line) is obtained at
incoherent summation of hard contributions.}\label{Fig7}
 }\hfill
\parbox[t]{0.47\textwidth}{
\vspace*{-.5\baselineskip}
\caption{\small The same that in Fig.\ref{Fig7} for $K^{-} p \to
\eta'\Lambda$.}\label{Fig8}   }
\end{figure}

Fortunately the uncertainty contained in (\ref{text32}) implies
negligibly small effect on the behavior of the differential
cross-sections. (This follows from the fact that in accordance with
(\ref{text28}), (\ref{text29}) the tensor contributions significantly
dominate even with $g_{K^{*} 0} = g_{K^{*}_2 0}$.) In fig.~\ref{Fig7}
and fig.~\ref{Fig8} we show data for the $\eta$ and $\eta'$ production
with the exponential fit for ``differential rate'' ${\mbox d}N/{\mbox
d}t$ \cite{GAMS}, and we plot theoretical curves (\ref{text28}) and
(\ref{text29}). At plotting the curves, we set the vertex factors
exponentially falling, but in the presence of nonexponential prefactors
we take somewhat greater the slopes compared with those in the
exponential fit in \cite{GAMS}. Simultaneously our slopes differ from
those predicted by Regge extrapolation. Namely, in the case of $\eta$
our slope is greater approximately on 30\%, and in the case of $\eta'$
is smaller approximately on 20\% compared with those obtained by Regge
extrapolation from the low-energy data. 

As follows from fig.~\ref{Fig7}, the differential cross-section for 
$\eta$ production does not contain any dip, and theoretical curve
(\ref{text28}) describes well this behavior. The difference between this
behavior and that at lower energies, we interpret as consequence of the
mode change of summation of hard contributions.

\section{Gluonium admixture}\label{sec4}

The above analysis is based on simplest scheme (\ref{text9}) for the
$\eta$--$\eta'$ mixing. However we can directly extend analysis to any
scheme of the mixing. Below we consider a generalization which includes
a gluonium state $|\,\eta^g>$ under assumption that $\eta$ does not
contain an admixture of $\eta^g$. This scheme requires an additional
mixing angle $\theta_G$. So, instead of (\ref{text9}) we have
\begin{eqnarray}\label{text33}
 |\,\eta \,> &=&
 \cos\theta \; |\,\eta^8> -  \sin\theta \; |\,\eta^0>\,,
 \nonumber\\ [-0.5\baselineskip]
 && \\ [-\baselineskip]
 \nonumber\\ [-0.5\baselineskip]
 |\,\eta'> &=&
 \cos\theta_G \sin\theta \; |\,\eta^8> +
 \cos\theta_G \cos\theta \; |\,\eta^0> +
 \sin\theta_G \; |\,\eta^g>\,.
 \nonumber
\end{eqnarray}
This scheme was first proposed as a solution to the axial Ward
identities for appropriate composite interpolating fields at the
requirement of the renormalization-group invariance of the pattern of
the mixing \cite{N2}. Later this scheme was repeatedly introduced on 
phenomenological basis
\cite{GAMS,Kou,KLOE1,KLOE2,EN,Thomas,E,Fleischer,Bigi,LHCb}.

With (\ref{text33}) the above formulas for the differential
cross-sections remain unchanged in the case of $\eta$ production, and
formulas for $\eta'$ undergo minimal changes. Namely, a common
factor $\cos^2 \theta_G$ appears in the r.h.s. in (\ref{text12}),
(\ref{text14}), (\ref{text23}), (\ref{text29}). The ratios of the
cross-sections are appropriately modified. In particular, (\ref{text16})
takes the form
\begin{equation}\label{text34}
 R_{\pi}^{\eta'/\eta}(0) \;  = \;
 \tan^2 (\theta + \theta_{id} - \delta) \cos^2 \theta_G \,.
\end{equation}
Correspondingly, with nonzero $\theta_G$ the value of $\theta$ must be
re-defined. At first glance, this means that many above results must
be revised.

We begin with considering the ratio $g_{K^{*} 0}/g_{K^{*}_2 0}$ in the
case of incoherent summation. Actually this ratio is of particular
interest since with $\theta_G=0$ its value was obtained close to the
boundary of solvability, see fig.~\ref{Fig6}. So with increasing
$|\theta_G|$ a solution may disappear at all. The latter situation would
mean a constraint for $\theta_G$.

In reality this happens. With increasing $\theta_G$ from zero, the
corridor of allowed values for $g_{K^{*} 0}/g_{K^{*}_2 0}$ in
fig.\ref{Fig6} goes rapidly to the left, while the corridor of allowed
values for $\theta$, obtained at equating (\ref{text34}) to
(\ref{text17}), slowly goes to the right. The corridors intersect so
long as $\cos^2 \theta_G \geq 0.93$. This gives 
\begin{equation}\label{text35}
 \sin^2 \theta_G \leq 0.07 \,,
\end{equation}
where the upper limit matches $g_{K^{*} 0}/g_{K^{*}_2 0} = 0$. 

Fortunately, with varying $\theta_G$ within (\ref{text35}) the 
corresponding value of $\theta$ is varying very little. For instance,
with $\sin^2 \theta_G = 0.07$, we have $\theta_p = (-17.5 \pm 1.2)^0$,
which coincides with\-in errors with (\ref{text18}). This means that our
above results numerically do not undergo noticeable changes. So the
qualitative results about the  mode of summation of hard contributions
remain unchanged.

It is worth comparing our estimate (\ref{text35}) with appropriate
results obtained in the same mixing scheme in other approaches. In
fact, (\ref{text35}) is compatible with
\cite{EN,Thomas,E,Fleischer,Bigi,LHCb}, and disagrees with
\cite{GAMS,KLOE1,KLOE2}. A disagreement with \cite{KLOE1,KLOE2} is
commented in \cite{EN,Thomas}. Especially we should comment a
disagreement between (\ref{text35}) and the estimate $\sin^2 \theta_G =
0.17 \pm 0.07$ of \cite{GAMS} because the latter estimate was obtained
not only in the same mixing scheme, but also on the basis of the same
data. Basically, there are two reasons for the disagreement. The first
one is that \cite{GAMS} in its theoretical analysis takes into
consideration only the diagrams of annihilation type, ignoring the
exchange-type diagrams, see fig.~\ref{Fig1}(a-d). The second reason is
that analysis of \cite{GAMS} is based on the model that does not take
into consideration the existence of two types of contributions,
conditioned by $K^*$ and $K^*_2$ trajectories.

\section{Discussion and conclusions}\label{sec5}

The above analysis discloses two essential features of the
charge-exchange reactions, which were not previously noted. Namely, they
go via the hard scattering of fast quarks, and soft interactions can
form a mode of summation of intermediate contributions. On this basis we
describe charge-exchan\-ge reactions in the combined approach which
joins together ideas of the parton model and Regge phenomenology, the
latter being used for the description of soft contributions. In the case
of coherent summation of hard contributions, the proposed approach in
its predictions is equivalent to conventional Regge approach. In the
case of incoherent summation, nontrivial differences appear.

The mine difference is a radical change of the behavior of the
differential cross-section $K^{-}p \to \eta \Lambda$. Namely, a
pronounced dip at $-t \approx$ 0.4--0.5 (GeV$\!$/c)$^2$, predicted in
the case of coherent summation, is replaced by a monotonic behavior in
the case of incoherent summation. Such a change has been observed
experimentally with a transition from relatively low to high momentum of
the $K^-$ beam. We interpret this phenomenon as a consequence of
restructuring the intermediate contributions because of capturing
uncorrelated partons arising due to the hard-scattering. In particular,
this leads to the mode change of summation of hard contributions.
Another manifestation of the mentioned restructuring is the appearance
of the dependence on energy in the vertex factors, which is forbidden in
the conventional Regge approach. We observe this effect in all reactions
under consideration. In the case of the charge-exchange reactions in
$\pi^{-}$ beams, we detect this effect at the confidence level of more
than 3$\sigma$.  

We carry out the main part of the calculations in the present work in
the framework of a simple mixing scheme which implies completeness of
two states, $\eta^0$ and $\eta^8$. In this scheme, we obtain estimation
(\ref{text18}) for the mixing. This result nearly coincides with that of
\cite{NICE1} but disagrees with that of \cite{Stanton}, both obtained on
the basis of the charge-exchange reactions. We explain the mentioned
disagreement as a consequence of not entirely correct method adopted in
\cite{Stanton} for the definition of the mixing angle. Further, we
consider a generalization of the mixing scheme allowing a gluonium
admixture in $\eta'$. We obtain a strong constraint on this admixture,
and we show that it has no noticeable effect on all other our results.
In particular, the corresponding change in the $\eta$--$\eta'$ mixing
does not go beyond the errors.

In a broad sense, the present work sheds light on the old problem of 
description at the microscopic level of the processes of exclusive
hadron scattering. (Currently, this problem has no satisfactory
solution.) In the case of the charge-exchange reactions we propose an
approach that combines the parton model and Regge phenomenology. In this
framework we explain the features of the behavior of differential
cross-sections $\pi^- p \to \eta (\eta') n$ and $K^- p \to \eta (\eta')
\Lambda$, which do not fit into representations of Regge approach. We
conclude that at the transition to higher energies in these reactions a
mode change of summation of intermediate contributions occurs, from
coherent mode to incoherent one. This result may have far-reaching
consequences for understanding details of various reactions, from
hadron-hadron to hadron-nucleus ones.

\bigskip
\noindent
{\small
{\bf Acknowledgements} \
The author is grateful to A.A.Godizov, V.A.Petrov and S.M.Troshin for
useful remarks, and to A.L.Kataev for interest to this work. Special
thanks to V.D.Samoy\-lenko for comments on experimental work
\cite{GAMS} and presented data in fig.~\ref{Fig7} and fig.~\ref{Fig8}.}

\end{document}